\documentclass[%
aps,
superscriptaddress,
showpacs,
prc,
twocolumn,
]{revtex4-1}

\usepackage{graphicx}% Include figure files
\usepackage{dcolumn}% Align table columns on decimal point
\usepackage{bm}% bold math
\begin{document}

\preprint{}

\title{Differential cross section and photon-beam asymmetry for 
the $\vec{\gamma} p$ $\rightarrow$ $\pi^{-}\Delta^{++}$(1232) reaction 
at forward $\pi^{-}$ angles for $E_{\gamma}$=1.5-2.95 GeV}% 

\newcommand{\KUADDRESS}{Department of Physics, Korea University, Seoul 02841, Republic of Korea}
\newcommand{\IBSADDRESS}{Rare Isotope Science Project, Institute for Basic Science, Daejeon 34047, Korea}
\newcommand{\OHIOADDRESS}{Department of Physics and Astronomy, Ohio University, Athens, Ohio 45701, USA}
\newcommand{\KYOTOADDRESS}{Department of Physics, Kyoto University, Kyoto 606-8502, Japan}
\newcommand{\CENTADDRESS}{Department of Physics, National Central University, Taoyuan City 32001, Taiwan}
\newcommand{\SINICAADDRESS}{Institute of Physics, Academia Sinica, Taipei 11529, Taiwan }
\newcommand{\CHEMADDRESS}{ChemMatCARS, The University of Chicago, Argonne, Illinois 60439, USA}
\newcommand{\NSRRCADDRESS}{Light Source Division, National Synchrotron Radiation Research Center, Hsinchu, 30076, Taiwan }
\newcommand{\RCNPADDRESS}{Research Center for Nuclear Physics, Osaka University, Ibaraki, Osaka 567-0047, Japan}
\newcommand{\JASRIADDRESS}{Japan Synchrotron Radiation Research Institute, Sayo, Hyogo 679-5143, Japan}
\newcommand{\NAGOYAADDRESS}{Kobayashi-Maskawa Institute, Nagoya University, Nagoya, Aichi 464-8602, Japan}
\newcommand{\TOHOKUADDRESS}{Research Center for Electron Photon Science, Tohoku University, Sendai, Miyagi 982-0826, Japan}
\newcommand{\OSAKAADDRESS}{Department of Physics, Osaka University, Toyonaka, Osaka 560-0043, Japan}
\newcommand{\TOKYOIADDRESS}{Department of Physics, Tokyo Institute of Technology, Tokyo 152-8551, Japan}
\newcommand{\GIFUADDRESS}{Department of Education, Gifu University, Gifu 501-1193, Japan}
\newcommand{\RIKENADDRESS}{RIKEN Nishina Center, 2-1 Hirosawa, Wako, Saitama 351-0198, Japan}
\newcommand{\JINRADDRESS}{Joint Institute for Nuclear Research, Dubna, Moscow Region, 142281, Russia}
\newcommand{\MSUADDRESS}{National Superconducting Cyclotron Laboratory, Michigan State University, East Lansing, MI 48824, USA}
\newcommand{\WAKAYAMAADDRESS}{Wakayama Medical College, Wakayama, 641-8509, Japan}
\newcommand{\SASKAADDRESS}{Department of Physics and Engineering Physics, University of Saskatchewan, Saskatoon, SK S7N 5E2, Canada}
\newcommand{\KEKADDRESS}{High Energy Accelerator Organization (KEK), Tsukuba, Ibaraki 305-0801, Japan}
\newcommand{\CONNEADDRESS}{Department of Physics, University of Connecticut, Storrs, CT 06269-3046, USA}
\newcommand{\FUKUIADDRESS}{Proton Therapy Center, Fukui Prefectural Hospital, Fukui 910-8526, Japan}
\newcommand{\JAEAADDRESS}{Advanced Science Research Center, 
Japan Atomic Energy Agency, Tokai, Ibaraki 319-1195, Japan}
\newcommand{\KRISSADDRESS}{Korea Research Institute of Standards 
and Science (KRISS), Daejeon 34113, Republic of Korea}
\newcommand{\PUKYONGADDRESS}{Department of Physics, Pukyong National University (PKNU), Busan 608-737, Republic of Korea}
\newcommand{\GENKENADDRESS}{National Institutes for Quantum and Radiological Science and Technology, Tokai, Ibaraki 319-1195, Japan}
\newcommand{\TOKYOADDRESS}{Department of Radiology, The University of Tokyo Hospital, Tokyo 113-8655, Japan}
\newcommand{\CASADDRESS}{Institute of High Energy Physics, Chinese Academy of Sciences, Beijing 100049, China}
\newcommand{\MICHIGANADDRESS}{Physics Department, University of Michigan, Michigan 48109-1040, USA}
\newcommand{\CROSSADDRESS}{Neutron Science and Technology Center, Comprehensive Research Organization for Science and Society (CROSS), Tokai, Ibaraki 319-1106, Japan}
\newcommand{\DUBNAADDRESS}{Joint Institute for Nuclear Research, Dubna, Moscow Region, 142281, Russia}

\author{H.~Kohri}\affiliation{\RCNPADDRESS}\affiliation{\SINICAADDRESS}
\author{S.~H.~Shiu}\affiliation{\SINICAADDRESS}\affiliation{\CENTADDRESS}
\author{W.~C.~Chang}\affiliation{\SINICAADDRESS}
\author{Y.~Yanai}\affiliation{\RCNPADDRESS}
\author{D.~S.~Ahn}\affiliation{\RIKENADDRESS}
\author{J.~K.~Ahn}\affiliation{\KUADDRESS}
\author{J.~Y.~Chen}\affiliation{\NSRRCADDRESS}
\author{S.~Dat$\acute{\rm{e}}$}\affiliation{\JASRIADDRESS}
\author{H.~Ejiri}\affiliation{\RCNPADDRESS}
\author{H.~Fujimura}\affiliation{\WAKAYAMAADDRESS}
\author{M.~Fujiwara}\affiliation{\RCNPADDRESS}\affiliation{\GENKENADDRESS}
\author{S.~Fukui}\affiliation{\RCNPADDRESS}
\author{W.~Gohn}\affiliation{\CONNEADDRESS}
\author{K.~Hicks}\affiliation{\OHIOADDRESS}
\author{A.~Hosaka}\affiliation{\RCNPADDRESS}
\author{T.~Hotta}\affiliation{\RCNPADDRESS}
\author{S.~H.~Hwang}\affiliation{\KRISSADDRESS}
\author{K.~Imai}\affiliation{\JAEAADDRESS}
\author{T.~Ishikawa}\affiliation{\TOHOKUADDRESS}
\author{K.~Joo}\affiliation{\CONNEADDRESS}
\author{Y.~Kato}\affiliation{\NAGOYAADDRESS}
\author{Y.~Kon}\affiliation{\RCNPADDRESS}
\author{H.~S.~Lee}\affiliation{\IBSADDRESS}
\author{Y.~Maeda}\affiliation{\FUKUIADDRESS}
\author{T.~Mibe}\affiliation{\KEKADDRESS}
\author{M.~Miyabe}\affiliation{\TOHOKUADDRESS}
\author{Y.~Morino}\affiliation{\KEKADDRESS}
\author{N.~Muramatsu}\affiliation{\TOHOKUADDRESS}
\author{T.~Nakano}\affiliation{\RCNPADDRESS}
\author{Y.~Nakatsugawa}\affiliation{\KEKADDRESS}\affiliation{\CASADDRESS}
\author{S.~i.~Nam}\affiliation{\PUKYONGADDRESS}
\author{M.~Niiyama}\affiliation{\KYOTOADDRESS}
\author{H.~Noumi}\affiliation{\RCNPADDRESS}
\author{Y.~Ohashi}\affiliation{\JASRIADDRESS}
\author{T.~Ohta}\affiliation{\RCNPADDRESS}\affiliation{\TOKYOADDRESS}
\author{M.~Oka}\affiliation{\RCNPADDRESS}
\author{J.~D.~Parker}\affiliation{\KYOTOADDRESS}\affiliation{\CROSSADDRESS}
\author{C.~Rangacharyulu}\affiliation{\SASKAADDRESS}
\author{S.~Y.~Ryu}\affiliation{\RCNPADDRESS}
\author{T.~Sawada}\affiliation{\SINICAADDRESS}\affiliation{\MICHIGANADDRESS}
\author{H.~Shimizu}\affiliation{\TOHOKUADDRESS}
\author{E.~A.~Strokovsky}\affiliation{\DUBNAADDRESS}\affiliation{\RCNPADDRESS}
\author{Y.~Sugaya}\affiliation{\RCNPADDRESS}
\author{M.~Sumihama}\affiliation{\GIFUADDRESS}
\author{T.~Tsunemi}\affiliation{\KYOTOADDRESS}
\author{M.~Uchida}\affiliation{\TOKYOIADDRESS}
\author{M.~Ungaro}\affiliation{\CONNEADDRESS}
\author{S.~Y.~Wang}\affiliation{\SINICAADDRESS}\affiliation{\CHEMADDRESS}
\author{M.~Yosoi}\affiliation{\RCNPADDRESS}

\collaboration{LEPS Collaboration}%\noaffiliation

\date{\today}% It is always \today, today,

\begin{abstract}
Differential cross sections and photon-beam asymmetries for the 
$\vec{\gamma} p$ $\rightarrow$ $\pi^{-}\Delta^{++}$(1232) 
reaction have been measured 
for 0.7$<$ $\cos\theta_{\pi}^{c.m.}$ $<$1 and $E_{\gamma}$=1.5-2.95 GeV 
at SPring-8/LEPS. 
The first-ever high statistics cross-section data are obtained 
in this kinematical region, and the asymmetry data 
for 1.5$<$ $\hspace*{-0.13cm}E_{\gamma}$(GeV)$<$2.8 are obtained 
for the first time. 
This reaction has a unique feature for 
studying the production mechanisms of a pure $u\bar{u}$ quark pair 
in the final state from the proton. 
Although there is no distinct peak structure in the cross sections, 
a non-negligible excess over the theoretical predictions is 
observed at $E_{\gamma}$=1.5-1.8 GeV. 
The asymmetries are found to be negative in most of 
the present kinematical regions, suggesting the dominance 
of $\pi$ exchange in the $t$ channel. 
The negative asymmetries at forward meson production angles are 
different from the asymmetries previously measured for 
the photoproduction reactions producing 
a $d\bar{d}$ or an $s\bar{s}$ quark pair in the final state. 
Advanced theoretical models introducing nucleon resonances and 
additional unnatural-parity exchanges are needed to 
reproduce the present data. 
\end{abstract}

%\pacs{13.60.Le, 13.60.Rj, 13.88.+e, 14.20.Jn, 25.20.Lj}

\maketitle

%\tableofcontents

%%%%%%% Introduction %%%%%%%%

%%%%%% ubar-u production in the final state
%The photoproduction of various mesons and baryons is important 
%to understand the production mechanisms of hadrons. 
The photoproduction of a $d\bar{d}$ quark pair and an $s\bar{s}$ 
quark pair in the final state has been extensively studied by 
the $\gamma p$ $\rightarrow$ $\pi^{+}n$~\cite{Kohri1,Dugger1,Dugger2} 
and $\gamma p$ $\rightarrow$ $K^{+}\Lambda$ and 
$K^{+}\Sigma^{0}$
~\cite{Sumihama,Kohri2,Lleres,Paterson,Shiu,Bradford,McCracken,Dey} 
reactions, respectively.
However, the production of a $u\bar{u}$ quark pair in the 
final state has not been well studied. 
Although the production of a $\pi^{0}$ meson, with a quark-model 
wavefunction of 
($u\bar{u}-d\bar{d}$)/$\sqrt{2}$, or $\eta$ meson, with a wavefunction 
of ($u\bar{u}$+$d\bar{d}$+$s\bar{s}$)/$\sqrt{3}$, includes 
the $u\bar{u}$ quark-pair production, an exclusive study 
of a pure $u\bar{u}$ quark-pair production is desired. 
The $\gamma p$ $\rightarrow$ $\pi^{-}\Delta^{++}$ reaction 
is a unique channel to study the 
photoproduction mechanism of a pure $u\bar{u}$ quark 
pair in the final state from the proton. 

%%%%%% pi- Delta++ differential cross sections
%Photoproduction of mesons is also of special importance to 
%search for missing nucleon resonances. 
In quark models, there exist more nucleon resonances than those 
experimentally observed so far~\cite{Capstick}. 
Since nucleon resonances have relatively wide widths and are 
overlapping each other, 
rich physics observables with a wide angular and energy range 
are needed to study new resonances. 
The differential cross sections for the 
$\gamma p$ $\rightarrow$ $\pi^{-}\Delta^{++}$ reaction 
were measured by SLAC at the higher energies of $E_{\gamma}$=4, 5, 
8, 11, and 16 GeV~\cite{Anderson,Boyarski,Boyarski2}. 
At medium energies, there are only scarce existing data taken 
by SLAC at 2.8 GeV~\cite{Ballam}, 
by CEA for $E_{\gamma}$=0.5-1.8 GeV~\cite{Cea}, 
by LAMP2 for $E_{\gamma}$=2.4-4.8 GeV~\cite{Barber}, 
by DESY for $E_{\gamma}$=0.3-5.8 GeV~\cite{Abbhhm}, 
and by SAPHIR for $E_{\gamma}$=1.1-2.6 GeV~\cite{Wu}. 
Although the $\pi^{-}\Delta^{++}$ final state is one of prospective 
channels to study new nucleon resonances~\cite{Capstick}, 
cross-section data with a  wide angular and energy range are 
missing in the world data set. 

%%%%% Photon-Beam Asymmetry
Basically, the photon-beam asymmetries are +1 for $\rho$ exchange and 
are $-$1 for $\pi$ exchange in the $t$ channel 
in the case of the $\vec{\gamma} p$ $\rightarrow$ $\pi^{-}\Delta^{++}$ 
reaction, which is the same as the case of the 
$\vec{\gamma} p$ $\rightarrow$ 
$\pi^{+}n$ reaction~\cite{Guidal}. 
There were three asymmetry measurements 
at the forward $\pi^{-}$ angles of $|t|<$0.5 GeV$^{2}$ 
(0.8$<\cos\theta^{c.m.}_{\pi}$) at 2.8 GeV, 4.7 GeV, 
and 16 GeV by SLAC~\cite{Ballam,Quinn}, where $t$ is the Mandelstam 
variable defined by $t$=($p_{\pi}-p_{\gamma}$)$^{2}$.  
Although negative asymmetries are suggested by these measurements, 
the number of the data points is limited and the data have 
large statistical uncertainties. 
In contrast, pseudoscalar meson photoproduction of either 
a $\pi^{+}$ or a $K^{+}$ has 
positive asymmetries at the forward meson angles of 
0.6$<\cos\theta^{c.m.}_{\pi,K}<$1 when the total energy $W$ is 
higher than the third nucleon resonance region 
$W\sim$1.7 GeV ($E_{\gamma}\sim$1.1 GeV) 
~\cite{Kohri1,Dugger2,Sumihama,Kohri2,Lleres,Paterson,Shiu}. 
The $\pi^{-}$ photoproduction data may well have a different reaction 
mechanism than that of other pseudoscalar mesons. 
Combining the $\pi^{-}\Delta^{++}$ data with the established 
$\pi^{+}$ and $K^{+}$ photoproduction data is helpful to 
achieve a unified understanding of hadron photoproduction. 
%, and 
%new asymmetry data are expected to play an important role in 
%searching for new nucleon resonances as studied in the 
%$K^{+}$ photoproduction~\cite{Mart}. 

%%%%%%%%%%%% End of Introduction 
In this Letter, we present the first-ever high statistics 
differential cross-section and photon-beam asymmetry data for the 
$\vec{\gamma} p$ $\rightarrow$ $\pi^{-}\Delta^{++}$ reaction 
at the forward $\pi^{-}$ angles of 0.7$<\cos\theta^{c.m.}_{\pi}<$1. 
The data obtained over the energy range of $E_{\gamma}$=1.5-2.95 GeV, 
covering most of the nucleon resonance region, 
enabled us to study both nucleon resonances and 
hadron photoproduction dynamics. 

%%%%%%%% Experiment %%%%%%%%%

The experiment was carried out using the LEPS beam line \cite{Nakano} 
at the SPring-8 facility in Japan. 
The photon beam was produced by the laser backscattering technique 
using a deep-UV laser with a wavelength of 257 nm~\cite{Muramatsu}. 
The energy range of tagged photons was from 1.5 to 2.96 GeV. 
The laser light was polarized linearly with a polarization 
degree of 98\%. 
The polarization of tagged photons was 88\% 
at 2.96 GeV and was 28\% at 1.5 GeV \cite{Hwang1}. 
The photon beam was incident on a liquid hydrogen target (LH$_{2}$) with 
a length of 16 cm. 

Charged particles emitted from the LH$_{2}$ target were detected at 
forward angles by using the LEPS spectrometer. 
The aerogel Cherenkov counter was not used, and 
electrons or positrons were vetoed using a 
plastic scintillation counter installed at the downstream position 
of the three drift chambers. 
For the details about the LEPS spectrometer, 
see Refs.~\cite{Nakano,Sumihama,Hwang1}. 

Events with a $\pi^{-}$ meson were identified from its mass within 
3$\sigma$ where $\sigma$ is the momentum dependent mass resolution and 
was measured to be 60 and 110 MeV/$c^{2}$ for 1 and 2 GeV/$c$ 
momentum pions, respectively. 
The events from the LH$_{2}$ target 
were selected by a cut on the $z$-vertex distribution. 
Contamination events from the start counter, placed downstream 
of the target, were 0.5\% at most. 

\begin{figure}[htb]
\includegraphics[height=4.1cm,width=7.6cm]{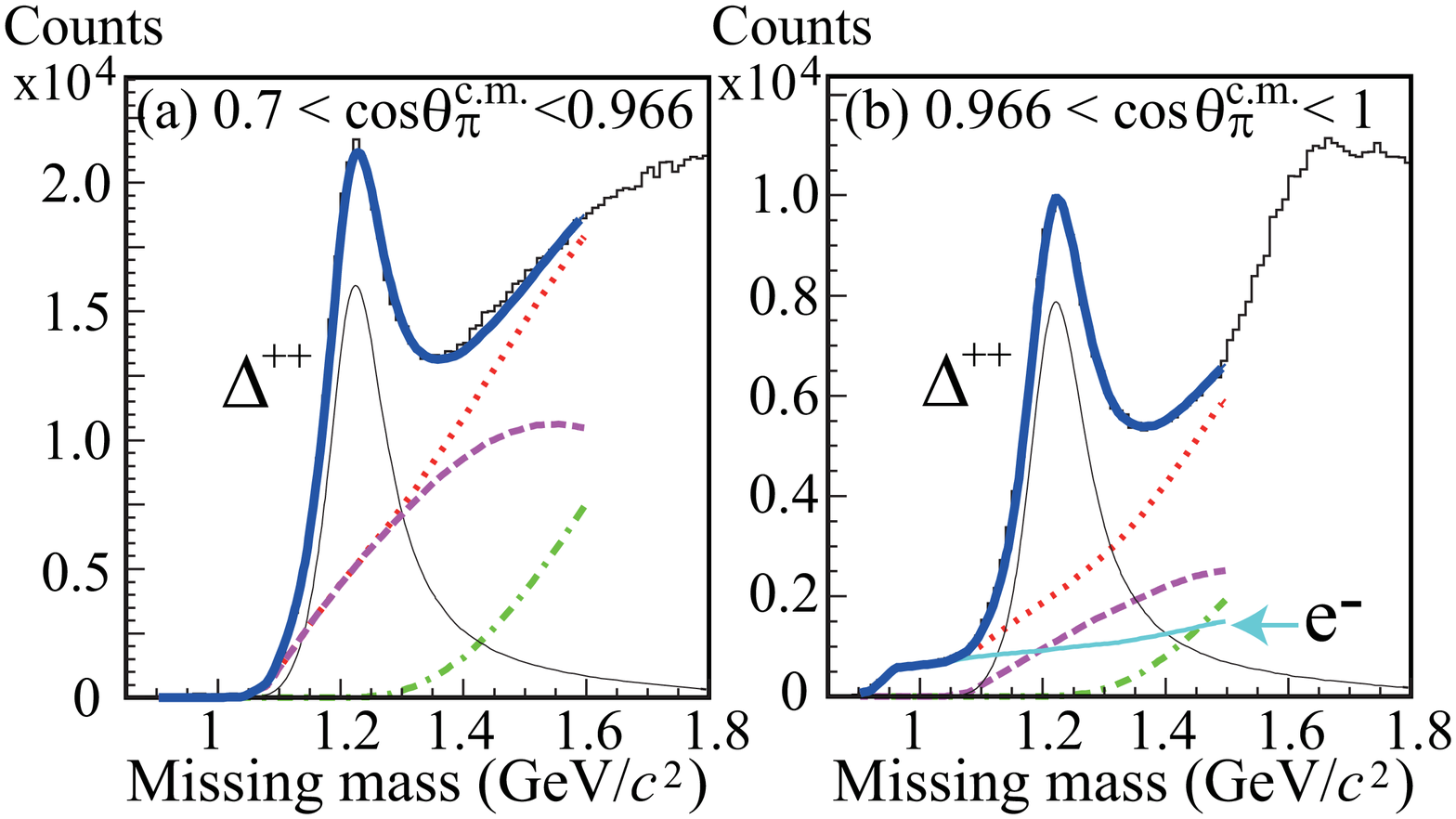}
\caption{\label{fig:miss} Missing-mass spectra for the 
$\gamma p$ $\rightarrow$ $\pi^{-}X$ reaction for 
$E_{\gamma}$=1.5-2.95 GeV. 
The thick solid curves (blue) are the results of the fits, and the thin 
solid curves (black) are the $\Delta^{++}$ contributions. 
The dotted curves (red) are the total contribution from backgrounds. 
The dashed (purple) and dotted-dashed (green) curves are the contributions 
from $\rho$/2$\pi$ and 3$\pi$ productions, respectively. 
The curve (light blue) indicated by the arrow is the contribution from the 
electron background.}
\end{figure}

Figure \ref{fig:miss} shows the missing-mass spectra for the 
$\gamma p$ $\rightarrow$ $\pi^{-}X$ reaction. 
The $\Delta^{++}$(1232) peaks are clearly observed at 1.23 GeV/$c^{2}$. 
The contribution from electrons, mainly originating from the $e^{+}e^{-}$ 
pair creation, is observed for 0.966$<\cos\theta_{\pi}^{c.m.}<$1. 
The number of $\pi^{-}\Delta^{++}$ events was about 400 k in total. 

%%%%%%%%%%%%%%%%%%%%% Figure 1 %%%%%%%%%%%%%%%%%%%%%%%

The $\gamma p$ $\rightarrow$ $\pi^{-}\Delta^{++}$ reaction events 
were selected by fitting a missing-mass spectrum with curves 
for the $\Delta^{++}$ peak, $\rho$, 2$\pi$, 3$\pi$ productions, 
and electron background based on GEANT simulations. 
For the $\rho$-meson productions, the differential cross sections 
and decay angular distributions in Ref.~\cite{Wu} were assumed, and the 
multi-pion productions according to Lorentz-invariant 
Fermi phase space were performed for the 2$\pi$ and 
3$\pi$ productions. 
The electron background events were generated to reproduce 
the momentum distributions of the real data. 
The number of adjustable parameters in the fit was 5 in total. 
No interference between the $\pi^{-}\Delta^{++}$ and 
other reactions was assumed in this analysis using single $\pi^{-}$ 
events. 
The shape of the $\Delta^{++}$(1232) was assumed to be given by 
a Jackson relativistic Breit-Wigner form~\cite{Jackson,Boyarski}, 
\begin{equation}
B(m)\propto \frac{m_{0}\Gamma(m)}{(m^{2}-m_{0}^{2})^{2}+m_{0}^{2}\Gamma^{2}(m)},
\end{equation}
with
\begin{equation}
\Gamma(m) = \Gamma(m_{0})\Bigl(\frac{q}{q_{0}}\Bigr)^{3}\Bigl(\frac{am_{\pi}^{2}+q_{0}^{2}}{am_{\pi}^{2}+q^{2}}\Bigr)\Bigl(\frac{m_{0}}{m}\Bigr),
\end{equation}
where $m_{0}$ = 1.232 GeV, $\Gamma(m_{0})$ = 0.117 GeV, 
%a = 2.2, $q$, $q_{0}$ are the c.m.-system momenta at masses $m$, $m_{0}$ 
%in the 
a = 2.2, with $q$($q_{0}$) being the c.m.-system momentum at masses 
$m$($m_{0}$) in the 
$\Delta^{++}$ rest system. 
As a result of the fit, the $\Delta^{++}$ yield for the relativistic 
Breit-Wigner form (including the tail) was obtained for 
each incident photon energy and angular bin. 
The acceptance of the LEPS spectrometer for 
$\pi^{-}$ mesons was obtained by the GEANT simulations. 
The differential cross sections for the $\pi^{-}\Delta^{++}$ reaction 
were obtained by using the same method described 
in Ref.~\cite{Sumihama}. 
The LEPS spectrometer has almost the same acceptance for 
the $\pi^{-}$ and $\pi^{+}$ mesons, and the cross sections for the 
$\gamma p$ $\rightarrow$ $\pi^{+}n$ reaction~\cite{Kohri1} obtained 
from the same data set agree well with the data obtained by 
CLAS~\cite{Dugger1} and DESY~\cite{Boschhorn2}. 

%%%%%%%%% Results %%%%%%%%%%%

%%%%%% Differential Cross Sections %%%%%%%%

\begin{figure}[htb]
\includegraphics[height=8.9cm,width=6.7cm]{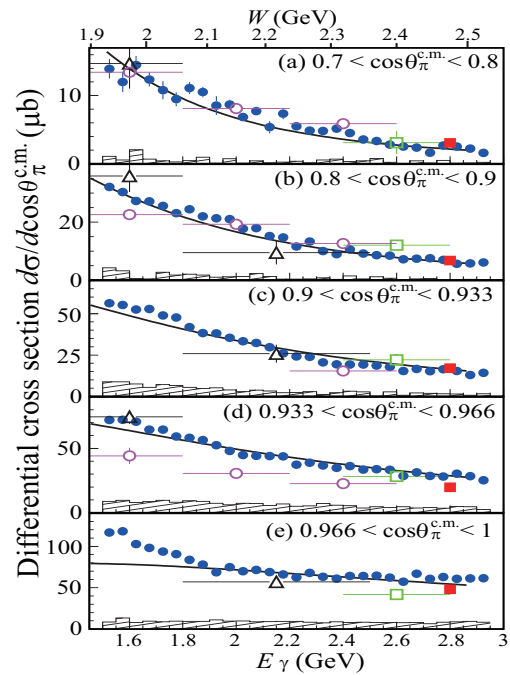}% 
\caption{\label{fig:cross} Differential cross sections 
for the $\gamma p$ $\rightarrow$ $\pi^{-}\Delta^{++}$ reaction as 
a function of $E_{\gamma}$. 
The closed circles, open circles, open triangles, open squares, and 
closed squares are the data obtained by LEPS, SAPHIR~\cite{Wu}, 
DESY~\cite{Abbhhm}, LAMP2~\cite{Barber}, and SLAC~\cite{Ballam}, respectively. 
Since the data obtained by the other groups had the form $d\sigma$/$dt$, 
they were transformed to the form $d\sigma$/$d\cos\theta^{c.m.}_{\pi}$ 
for the comparison. 
The hatched histograms are the systematic uncertainties 
due to the selection of the $\Delta^{++}$ shape. 
The solid curves are the results of theoretical calculations 
by S.~i. Nam~\cite{Nam}.}
\end{figure}

Figure \ref{fig:cross} shows the differential cross sections 
for the $\gamma p$ $\rightarrow$ $\pi^{-}\Delta^{++}$ reaction 
as a function of $E_{\gamma}$. 
The cross sections decrease rapidly with increasing photon energy
for 0.7$<\cos\theta_{\pi}^{c.m.}<$0.933. 
The energy dependence of the cross sections 
is small for 0.966$<\cos\theta_{\pi}^{c.m.}<$1. 
There is no distinct peak structure in the cross sections. 
The cross sections increase rapidly when the $\pi^{-}$ angle 
becomes smaller. 
A strong forward peaking of the cross sections is observed. 
Similar strong forward peaking at $|t|<$0.2 GeV$^{2}$ 
was reported for $E_{\gamma}$=5, 
8, 11, and 16 GeV by SLAC~\cite{Boyarski}. 
The momentum transfer of $|t|<$0.2 GeV$^{2}$ corresponds to 
the $\pi^{-}$ angular region of 0.9$<\cos\theta^{c.m.}_{\pi}<$1 
in the present experiment. 
The exchange of an isospin $I=1$ meson ($\pi$ or $\rho$) in the $t$ channel 
is expected to be the dominant reaction mechanism in the present 
kinematical region. 
Since the $\rho$-meson exchange contribution becomes weak 
at forward $\pi$ angles in pion photoproduction~\cite{Guidal}, 
$\pi$-meson exchange is inferred to 
play an important role in making the forward-peaking $\pi^{-}\Delta^{++}$ 
cross sections for 0.9$<\cos\theta_{\pi}^{c.m.}<$1. 

%%%%%%%%%%%%%%%%%%%%% Figure 2 %%%%%%%%%%%%%%%%%%%%%%%

The LEPS cross sections for the $\pi^{-}\Delta^{++}$ reaction are 
in good agreement with the cross sections measured by 
DESY~\cite{Abbhhm} and SLAC~\cite{Ballam} overall. 
The LEPS cross sections also agree well with those measured 
by LAMP2~\cite{Barber} except for 
0.966$<$ $\cos\theta_{\pi}^{c.m.}<$1. 
The cross sections by SAPHIR~\cite{Wu} agree with the LEPS data 
for 0.7$<$ $\cos\theta_{\pi}^{c.m.}<$0.933 and are 
smaller than the LEPS data for 
0.933$<$ $\cos\theta_{\pi}^{c.m.}<$0.966. 
Since the $\pi^{-}\Delta^{++}$ reaction has strong forward-peaking 
cross sections, small differences in the $\pi^{-}$ angular 
regions between the SAPHIR and present data 
might cause these disagreements. 

Theoretical calculations, employing the tree-level Regge-Born 
interpolation model without nucleon resonance contributions 
by S.~i.~Nam~\cite{Nam}, almost reproduce the present cross sections. 
Although the cutoff mass parameter was optimized from 450 MeV to 500 MeV 
to fit the data, the energy dependence of the cross sections 
for 0.9$<\cos\theta^{c.m.}_{\pi}<$1 and $E_{\gamma}$=1.5-1.8 GeV 
was not reproduced. 
One of the possible explanations for this discrepancy can be 
attributed to the absence of resonance contributions 
in the theory. 
For instance, $N^{*}$(1900, 3/2$^{+}$), which strongly couples 
to $\pi\Delta$, could be responsible for describing a bump observed 
in the cross section data. 
Since the $s$-channel structures observed in the SAPHIR total 
cross sections~\cite{Wu} seem to continue up to 
$E_{\gamma}\sim$2 GeV, the excess in the present 
cross sections might be the tail of $s$-channel 
structures. 

In this analysis, the relativistic Breit-Wigner form in Eq.~(1) 
used by SLAC~\cite{Boyarski} was employed. 
Another analysis with a different relativistic Breit-Wigner shape 
used by DESY~\cite{Abbhhm} 
gave smaller cross sections than the present cross sections. 
The differences between the cross sections obtained by 
the two Breit-Wigner forms are 10\% on average and 
are shown in Fig.~\ref{fig:cross} as 
the largest systematic uncertainties. 
Since both of the relativistic Breit-Wigner shapes 
originated from the shapes studied by Jackson~\cite{Jackson}, 
the differences in shape are not so large. 
New analyses using different shapes for the $\rho$, 2$\pi$, and 
3$\pi$ production events generated by the simulations with 
different momentum and angular distributions were performed. 
The differences between the original and new cross sections 
were smaller than 10\% for most of the data points. 
Systematic uncertainties of target thickness and photon flux 
are 1\% and 3\%, respectively.

%%%%%%%%%% Photon-beam Asymmetry %%%%%%%%%

\begin{figure}[htb]
\includegraphics[height=4.8cm,width=7.4cm]{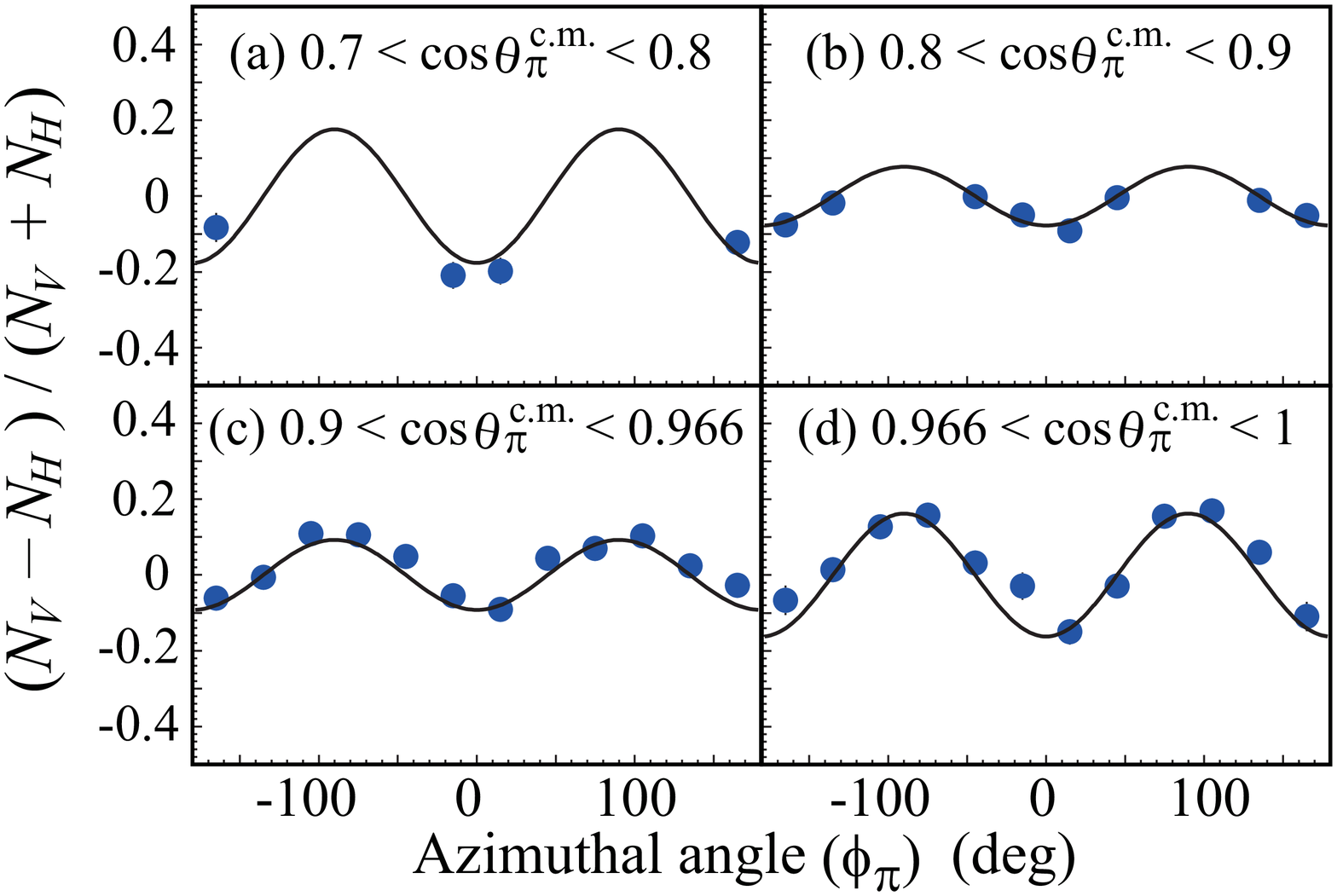}%
\caption{\label{fig:azimuthalangle} The ratio 
($N_{V}-N_{H}$)/($N_{V}$+$N_{H}$) as a function 
of $\pi^{-}$ azimuthal 
angle ($\phi_{\pi}$) for the $\vec{\gamma} p$ $\rightarrow$ 
$\pi^{-}\Delta^{++}$ reaction for $E_{\gamma}$=1.5-2.9 GeV. 
The curves are the result of the fits with 
$P_{\gamma}\Sigma \cos2\phi_{\pi}$. }
\end{figure}

The $\vec{\gamma}p$ $\rightarrow$ $\pi^{-}\Delta^{++}$ reaction 
data were measured using vertically and horizontally polarized photons. 
The photon-beam asymmetry $\Sigma$ is given as 
\begin{equation}
P_{\gamma}\Sigma \cos2\phi_{\pi} = \frac{N_{V} - N_{H}}{N_{V} + N_{H}}, 
\end{equation}
where $N_{V}$ and $N_{H}$ are the $\pi^{-}\Delta^{++}$ yields with 
vertically and horizontally polarized photons, respectively, 
after correcting for the difference of photon flux in both polarizations. 
$P_{\gamma}$ is the polarization of the photons and 
$\phi_{\pi}$ is the $\pi^{-}$ azimuthal angle. 
The $\pi^{-}\Delta^{++}$ yield is obtained by fitting a 
missing-mass spectrum for each $\phi_{\pi}$, $\cos\theta^{c.m.}_{\pi}$, 
and $E_{\gamma}$ region. 
Figure \ref{fig:azimuthalangle} shows the ratio 
($N_{V}-N_{H}$)/($N_{V}$+$N_{H}$) for the 
$\pi^{-}\Delta^{++}$ reaction events for $E_{\gamma}$=1.5-2.9 GeV. 

%%%%%%%%%%%%%%%%%%%%% Figure 3 %%%%%%%%%%%%%%%%%%%%%%%

Since the LEPS spectrometer has a wide acceptance for the horizontal 
direction and a narrow acceptance for the vertical direction, 
the number of events is small at around $\phi_{\pi}$=$\pm$90$^{\circ}$ 
for 0.7$<$ $\cos\theta_{\pi}^{c.m.}$ $<$0.9. 
The ratio ($N_{V}-N_{H}$)/($N_{V}$+$N_{H}$) is 
large at $\pm$90$^{\circ}$ and 
is small at 0$^{\circ}$ and 180$^{\circ}$. 
The $\pi^{-}$ mesons prefer to scatter at $\phi_{\pi}$ angles parallel 
to the polarization plane. 
The photon-beam asymmetries are therefore negative. 

\begin{figure}[htb]
\includegraphics[height=8.9cm,width=6.7cm]{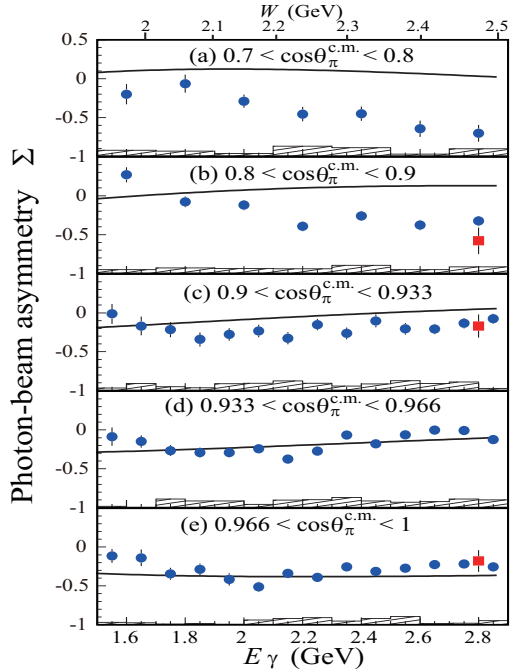}% 
\caption{\label{fig:asymmetry} Photon-beam asymmetries for 
the $\vec{\gamma} p$ $\rightarrow$ $\pi^{-}\Delta^{++}$ reaction 
as a function of $E_{\gamma}$. 
The circles and squares are the data obtained by LEPS and 
SLAC~\cite{Ballam}, respectively. 
SLAC measured three asymmetries at $|t|$=0.2-0.5, 0.1-0.2, and 
$|t_{min}|$-0.1 GeV$^{2}$, and are plotted as squares 
in (b), (c), and (e). 
The hatched histograms are the systematic uncertainties 
due to the selection of the $\Delta^{++}$ shape. 
The solid curves are the results of theoretical calculations by 
S.~i.~Nam~\cite{Nam}. }
\end{figure}

Figure \ref{fig:asymmetry} shows the photon-beam asymmetries for the 
$\vec{\gamma} p$ $\rightarrow$ $\pi^{-}\Delta^{++}$ reaction. 
The systematic uncertainty of the laser polarization is 
$\delta\Sigma$=0.02. 
The effect of the electron contamination in the $\pi^{-}$ sample 
is removed and that of the start counter contamination 
in the LH$_{2}$ target selection is negligibly small. 
The limited number of bins for the $\pi^{-}$ azimuthal angle 
in Fig.~\ref{fig:azimuthalangle} 
reduces absolute asymmetry values by 7\% on average. 
The asymmetries obtained using a different relativistic 
Breit-Wigner shape~\cite{Abbhhm} agree with the present 
asymmetries. 
The differences between the two asymmetries are 
$\delta\Sigma$=0.07 on average and are shown in Fig.~\ref{fig:asymmetry} 
as the largest systematic uncertainties. 
New analyses using different shapes for the $\rho$, 2$\pi$, and 3$\pi$ 
production events generated by the simulations with different 
momentum and angular distributions were performed. 
The differences between the original and new asymmetries were 
smaller than the statistical errors. 
For the confirmation of the correctness of the asymmetries, 
the sideband subtraction analysis using the sideband events of 
the $\Delta^{++}$ peak was performed, and the result of this analysis 
well reproduced the asymmetries in Fig.~\ref{fig:asymmetry}.

%%%%%%%%%%%%%%%%%%%%% Figure 4 %%%%%%%%%%%%%%%%%%%%%%%

The asymmetries are found to be negative in most of the LEPS 
kinematical region, which may be explained by $\pi$-meson exchange 
in the $t$ channel. 
The same interpretation is obtained from the strong forward 
peaking of the cross sections observed in Fig~\ref{fig:cross}. 
We have observed positive asymmetries at forward pseudoscalar 
meson angles in most $q\bar{q}$ productions in the final 
state from the proton, 
such as a $d\bar{d}$ production with $\vec{\gamma} p$ $\rightarrow$ 
$\pi^{+}n$~\cite{Kohri1,Dugger2} and an $s\bar{s}$ production with 
$\vec{\gamma} p$ $\rightarrow$ $K^{+}\Lambda$ and 
$K^{+}\Sigma^{0}$~\cite{Sumihama,Shiu,Kohri2,Lleres,Paterson}. 
In addition, the photoproduction reactions of neutral pseudoscalar mesons, 
such as $\vec{\gamma} p$ $\rightarrow$ 
$\pi^{0}p$~\cite{Sparks,Dugger2} 
and $\eta p$~\cite{Elsner,Collins}, also have positive 
asymmetries at forward meson production angles. 
It is quite interesting that only pure $u\bar{u}$ 
production in the final state has negative asymmetries. 
Since preliminary results of the 
$\vec{\gamma} p$ $\rightarrow$ $\pi^{+}\Delta^{0}$ reaction 
show positive asymmetries~\cite{Kohri3}, the production 
of the spin-parity 3/2$^{+}$ baryon does not necessarily cause 
the negative asymmetries. 

SLAC measured three asymmetries at $E_{\gamma}$=2.8 GeV~\cite{Ballam}. 
The agreement between the LEPS and SLAC data is reasonable 
for 0.9$<\cos\theta_{\pi}^{c.m.}<$0.933 and 
0.966$<\cos\theta_{\pi}^{c.m.}<$1. 
The SLAC data for 0.8$<\cos\theta_{\pi}^{c.m.}<$0.9 
is slightly smaller than for the LEPS data. 

Since the asymmetry calculated in Ref.~\cite{Nam} has an opposite 
definition, 
it is corrected to meet the present definition in Eq. (3). 
Theoretical calculations by S.~i.~Nam~\cite{Nam} almost reproduce 
the negative asymmetry data for 0.933$<\cos\theta^{c.m.}_{\pi}<$1. 
As the $\pi$ angle becomes larger, the calculations predict 
small positive asymmetries since the $\pi$-exchange contribution 
becomes small. 
The inconsistency between the data and the calculations 
becomes large for 0.7$<\cos\theta^{c.m.}_{\pi}<$0.9. 
This inconsistency is inferred to be due to the possible existence 
of a small but finite additional unnatural-parity exchange contribution 
%such as the axial-vector meson $b_1$(1235), 
not taken into account in the theory calculations. 
Smaller absolute asymmetry values at $E_{\gamma}$=1.5-1.7 GeV 
and 0.9$<\cos\theta^{c.m.}_{\pi}<$1 might be caused 
by the bump observed in the cross sections (Fig.~\ref{fig:cross}). 

%%%%%% Summary %%%%%%%

In summary, we have carried out a photoproduction experiment observing 
the $\vec{\gamma} p$ $\rightarrow$ $\pi^{-}\Delta^{++}$ reaction 
by using linearly 
polarized tagged photons with energies from 1.5 to 2.95 GeV. 
Differential cross sections and photon-beam asymmetries 
have been measured for 0.7$<\cos\theta_{\pi}^{c.m.}<$1. 
There is no distinct peak structure in the cross sections. 
However, a non-negligible excess of the cross sections, possibly due to 
the tail of nucleon or $\Delta$ resonances, over the theoretical 
predictions is observed at $E_{\gamma}$=1.5-1.8 GeV. 
Strong forward-peaking cross sections, expected from  
$\pi$ exchange in the $t$ channel, are observed. 
The asymmetries for the $\pi^{-}\Delta^{++}$ reaction 
are found to be negative in most of the present kinematical regions, 
which suggests that the $\pi$ exchange in the $t$ channel is dominant. 
The negative asymmetries are unusual in the photoproduction 
reactions from the proton studied 
in the 
past~\cite{Kohri1,Dugger2,Sumihama,Shiu,Kohri2,Lleres,Paterson}. 
Analogous results were obtained in the measurements of 
the single-spin asymmetries for the $pp$ or $ep$ 
reactions~\cite{Aidala}, where 
inclusive $\pi^{-}$ production has negative asymmetries while 
inclusive $\pi^{+}$ and $K^{+}$ productions have positive asymmetries. 
The $\pi^{-}$ production is inferred to have a different reaction 
mechanism from the $\pi^{+}$ and $K^{+}$ productions. 
%Although these measurements were performed at higher energies 
%than the present ones, analogous results of unusual $\pi^{-}$ 
%productions might be a hint to largely advance our understanding 
%of meson productions over a wide energy range. 
% NOTE: while the above is true, it distracts the reader from the main point.
The $\gamma p$ $\rightarrow$ $\pi^{-}\Delta^{++}$ reaction data 
provide a unique chance 
for studying the $u\bar{u}$ quark pair production. 
The combination of these data with the established data for the 
$d\bar{d}$ and $s\bar{s}$ quark pair productions is helpful to 
achieve unified understanding of the hadron photoproduction. 

\begin{acknowledgments}
The authors gratefully acknowledge the staff of 
the SPring-8 facility for the supports with excellent 
experimental conditions. 
The experiments were performed at the BL33LEP of SPring-8 
with the approval of the Japan Synchrotron Radiation Research 
Institute (JASRI) as a contract beamline (Proposal No. BL33LEP/6001). 
H.K. thanks Prof. E. Oset, Prof. T. Mart, Prof. S. Kumano, and 
Prof. H. Kamano for fruitful discussions. 
This research was supported in part by 
the Ministry of Education, Science, Sports and Culture of Japan, 
the National Science Council of the Republic of China, 
the National Research Foundation of Korea, and 
the U.S. National Science Foundation. 
\end{acknowledgments}

\nocite{*}

\bibliography{apssamp}% Produces the bibliography via BibTeX.

\end{document}